# Unusual suppression of superconductivity in YNi$_2$B$_2$C under neutron irradiation.

A.E. Karkin[*], Yu.N. Akshentsev, B.N. Goshchitskii

*Institute of Metal Physics, Ural Branch, RAS, Ekaterinburg 620219, ul. S. Kovalevskoi 18*
[*]*e-mail: aekarkin@rambler.ru*

The behavior of electrical resistivity $\rho(T)$, temperature of superconducting transition $T_c$, and upper critical field $H_{c2}(T)$ of polycrystalline YNi$_2$B$_2$C after irradiation with thermal neutron and subsequent high-temperature isochronous annealings in the temperature range $T_{ann}$ = 100 – 1000°C has been studied. It is found that irradiation of YNi$_2$B$_2$C with a fluence of $1*10^{19}$ cm$^{-2}$ results in a suppression of superconductivity. The disordered state obtained is reversible, i.e., the initial values of $\rho(T)$, $T_c$, and $H_{c2}(T)$ become virtually restored after annealing at temperatures up to $T_{ann}$ = 1000°C. For a sample in the superconducting state ($T_c$ = 5.5 – 14.5 K) there is observed a squared dependence $\rho(T) = \rho_0 + a_2 T^2$, the coefficient $a_2$ (proportional to the squared electronic mass $m^*$) being approximately constant. The form of dependence of $T_c$ on $\rho_0$ can be accounted for by the suppression of two superconducting gaps $\Delta_1$ and $\Delta_2$, $\Delta_1 \sim 2\Delta_2$; the rate of degradation of $\Delta_1$ being ~3 times as high as $\Delta_2$. The dependences of $dH_{c2}/dT$ on $\rho_0$ and $T_c$ can be described using relationships for a superconductor in an intermediate limit (the coherence length $\xi_0$ is on the order of free pass for an electron $l_{tr}$) under assumption of an approximately constant density of electronic states at the Fermi level $N(E_F)$. The observed behavior of $T_c$ upon irradiation of YNi$_2$B$_2$C obviously contradicts the common conception on the pure electron-phonon mechanism of superconductivity in compounds of this type, suggesting an anomalous type of superconducting pairing.

Superconducting (SC) compounds of the RNi$_2$B$_2$C type (R=Y, Lu, Er, Ho…) attracts much attention in view of their relatively high temperature of the SC transition $T_c$ ~15 K and different effects connected with the coexistence of magnetic and SC types of ordering [1, 2]. High $T_c$ values for compounds of this type are most often associated with high density of electronic states at the Fermi level $N(E_F)$ and, correspondingly, coefficient of heat capacity $\gamma \approx 18$ (mJ)/(mole·K$^2$) [3, 4] within the concept of conventional isotropic pairing of the s-type [4, 5, 6, 7, 8] with strong electron-phonon interaction but with two SC gaps $\Delta_1$ = 2.67 meV and $\Delta_2$ = 1.19 meV [9]. Yet, majority of experimental facts poorly agree with such a scenario, and at present, it is unclear whether such superconductors are conventional or exotic [9].

Useful information on the type of pairing can be gained while studying effects of scattering at impurities (defects). For conventional one-band s-wave superconductors with the electron-phonon interaction a nonmagnetic scattering, unlike the magnetic one, does not lead to depairing (Anderson's theorem); the superconducting transition temperature $T_c$ does not change in the case of an isotropic one-band superconductor. For a more complicated band structure, nonmagnetic scattering can lead to different effects of "averaging" of the superconducting gap function $\Delta(E)$, density of states $N(E)$, and other spectral functions, which may result in a decrease (in more seldom cases, increase) of $T_c$ [10, 11, 12]. However, if $N(E_F)$ does not vanish, $T_c$ remains finite. On contrary, the nonmagnetic scattering leads to depairing in the case of unconventional superconductors for which the part of quasiparticles leading to the Cooper pairing of electrons is mediated by magnetic excitations. A necessary condition for the existence of such mechanism of superconductivity is different signs of the gap function $\Delta$ for different parts of the Fermi surface, so that the effects of "averaging", arising upon a relatively weak nonmagnetic scattering, result in $\Delta = 0$, and consequently, complete suppression of superconductivity.

Experiments on substitution by magnetic impurities (Y by Gd or Ni by Co) demonstrated a fast decrease of $T_c$ [13, 14], which as if agrees with Abrikosov-Gor'kov model (AG) describing effects of depairing upon magnetic scattering [15]. However, measurements of elec-tronic heat capacity showed that, for example, in the Y(Ni$_{2-x}$Co$_x$)$_2$B$_2$C system there is observed a very strong (by a factor of 2.5) decrease of $\gamma$ [16], which means that the effects of doping, leading to a shift of the Fermi level and decrease of $N(E_F)$, play a more important role than the effects of magnetic scattering. Actually, isovalent substitutions in the Y(Ni$_{1-x}$Pt$_x$)$_2$B$_2$C system result in a slower decrease of $T_c$ [17].

In this work we studied the influence of radiation defects induced by neutron irradiation on the properties of normal and SC states of polycrystalline samples of YNi$_2$B$_2$C. In [12] it is shown that radiation defects are ideal centers of nonmagnetic scattering and, therefore, irradiation is one of the testing experiments on the determination of the type of superconducting pairing.

Polycrystalline samples of YNi$_2$B$_2$C were prepared by zone melting of stoichiometric amounts of initial elements. The samples contained impurities of several minor phases whose structure was not identified. Below, we present the results of investigation of temperature dependences of electrical resistivity $\rho(T)$, temperature of superconducting transition $T_c$, and upper critical field $H_{c2}(T)$ for a sample of YNi$_2$B$_2$C 3.5·1.7·0.7 mm$^3$ in dimensions after irradiation with the thermal-neutron fluence 1·10$^{19}$ cm$^{-2}$ at the temperature $T_{irr} \approx 50°$C and subsequent high-temperature isochronous annealings in the temperature range $T_{ann} = 100 – 1000°$C.

Radiation-induced distortions of the crystalline structure of YNi$_2$B$_2$C upon irradiation in the nuclear reactor are given rise by nuclear interactions with both fast and thermal neutrons according to the nuclear reaction:

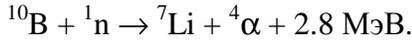
$^{10}$B + $^1$n → $^7$Li + $^4\alpha$ + 2.8 МэВ.

Since the cross-section of interaction of the thermal neutrons with a natural mixture of boron nuclei ($\Sigma \approx 7.5·10^{-22}$ см$^2$) is approximately two orders as high as typical values for the elastic nuclear interaction with fast neutrons ($\Sigma \approx 5·10^{-24}$ cm$^2$) and so is the energy value of an $\alpha$-particle (2.8 MeV) compared to the energy of an atom primarily knocked out by fast neutrons (tens KeV), the distortions of the crystalline structure generated by thermal neutrons turn out 4 orders of magnitude as effective as those in the case of fast neutrons. Therefore, in the case under consideration fast–neutron-related effects can be neglected.

The free path of a neutron in YNi$_2$B$_2$C is of the same order as the sample thickness (0.5 and 0.7 mm, respectively). However, because of multiple atomic displacements that arise upon irradiation with the fluence 1·10$^{19}$ cm$^{-2}$ one can expect a uniform distribution of radiation defects over the sample volume with a maximal concentration for a given irradiation temperature $T_{irr} \approx 50°$C.

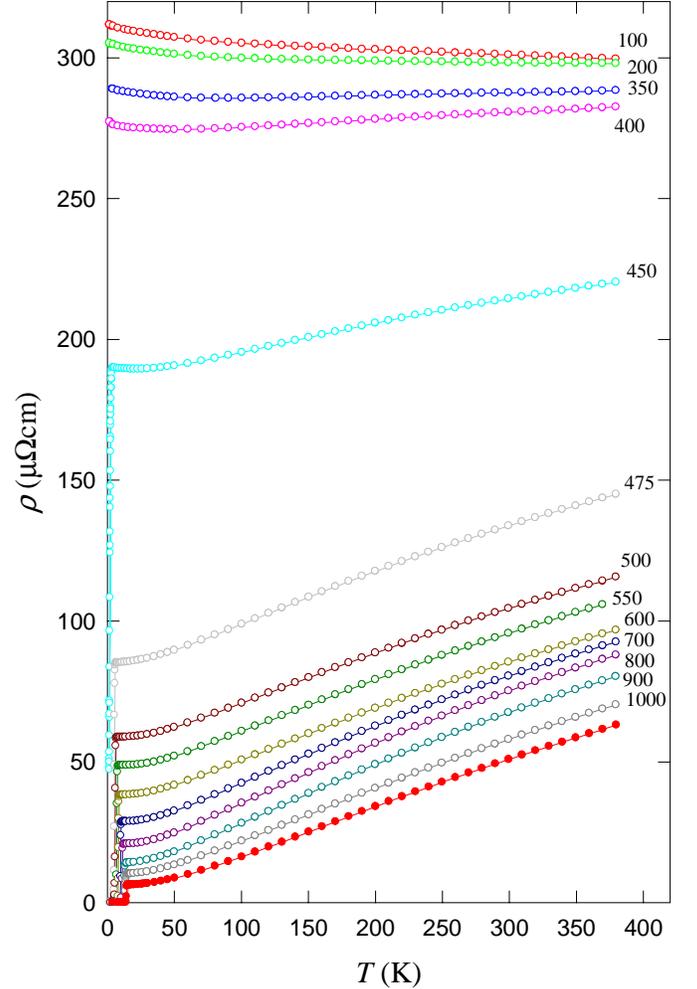

Fig. 1. Temperature dependences of resistivity $\rho(T)$ for the initial sample of YNi$_2$B$_2$C (●); after irradiation with the fluence 1·10$^{19}$ cm$^{-2}$ and subsequent annealings (○); figures show the annealing temperatures in °C.

In Fig. 1 there are shown temperature dependences of resistivity $\rho(T)$ for the YNi$_2$B$_2$C sample in the initial state and after irradiation and annealings. Since band calculations [18, 19] show a three-dimensional structure, which is supported by a relatively weak anisotropy of $\rho(T)$ [20] and $H_{c2}(T)$ [21] one can expect the absolute magnitude of the resistivity for the polycrystalline samples to be nearly the same as for single crystals. Indeed, the $\rho$ value for our sample at room temperature is equal to ~45 μΩcm, which slightly exceeds typical values for single crystals ~35 μΩcm [20].

The irradiation of YNi$_2$B$_2$C results in the complete suppression of the SC state and a significant transformation of the $\rho(T)$ dependences. As is seen in Fig. 1, after irradiation, $d\rho/dT < 0$ at $T \leq 380$ K. In the course of annealing, the temperature range in which $d\rho/dT < 0$, narrows and after annealing at $T_{ann} = 450^\circ$C is retained only at low temperatures when superconductivity restores itself for the first time. A similar behavior of $\rho(T)$ was observed for many strongly disordered compounds when the free path $l_{tr}$ for electrons is close to the lattice parameter $a_0$ [10]. At $T_{ann} > 450^\circ$C, in the temperature range $T \leq 70$ K there arises an approximately quadratic dependence $\rho(T) = \rho_0 + a_2 T^2$ with a nearly constant value $a_2 \approx 1.5 \cdot 10^{-3}$ μΩcm/K$^2$ (Fig. 2), whereas $\rho(T)$ and $T_c$ take up virtually their initial values after annealing at $T_{ann} = 1000^\circ$C.

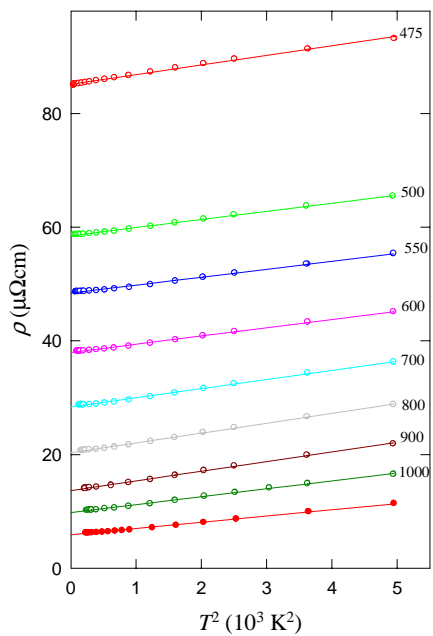

Fig. 2. $\rho(T)$ as a function of $T^2$ for the YNi$_2$B$_2$C sample; curve notations are the same as in Fig. 1.

The average value $a_2 \approx 1.5 \cdot 10^{-3}$ μΩcm/K$^2$ agrees rather well with the value $a_2 = 1.8 \cdot 10^{-3}$ μΩcm/K$^2$ for LuNi$_2$B$_2$C [21]. In describing the experimental data with the power-law dependence $\rho(T) = \rho_0 + a_n T^n$, we obtain n = 2.25, $a_n = 0.4 \cdot 10^{-3}$ μΩcm/K$^n$, which agree with the data for YNi$_2$B$_2$C as well [21]. The quadratic dependence of $\rho(T)$ observed at low temperatures is commonly ascribed to the electron-electron scattering, $a_2$ being proportional to either squared electron mass $m^*$ or density of states $N(E_F)$. For the initial sample, the ratio $a_2/\gamma^2 \approx 0.5 \cdot 10^{-5}$, where $a_2$ is in units of μΩcm, and $\gamma$ is in units of (mJ)/(mole·K$^2$), which agrees quite well with the empirical relationship of Kadowaki-Woods [22] $a_2/\gamma^2 = 1 \cdot 10^{-5}$. Hence, the weak change of $a_2$ for YNi$_2$B$_2$C in the SC state ($T_c = 5.5 - 14.5$ K) indicates that the density of states $N(E_F)$ does not change.

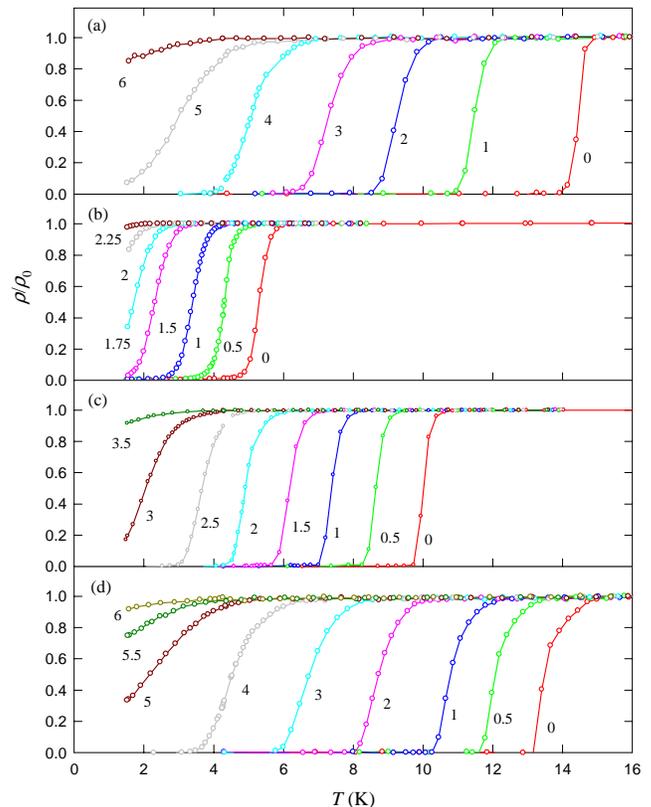

Fig. 3. Resistivity curves of the YNi$_2$B$_2$C sample upon the SC transition: in the initial state (a), after irradiation and annealing at temperatures 475 (b), 700 (c) and 1000$^\circ$C (d); figures denote values of magnetic field $H$ in T.

The resistivity curves of the SC transition (Fig. 3) show a notable widening of the temperature range in magnetic fields $H > 3$ Tл. The values of $H_{c2}(T)$ determined, in accordance criterion as 0.9 of the value of residual resistivity $\rho_0$, are shown in Fig. 4. The application of other criteria such as, for example, 0.5 or 0.95 yields somewhat smaller or larger, respectively, slope of the curves $H_{c2}(T)$ with similar temperature dependences. These curves are significantly declined from the theoretical curve by the Helfand-Werthamer (HW) model [23]: in the vicinity of $T_c$ there is observed a remarkable positive curvature of $H_{c2}$, whereas in the low-temperature range, a negative curvature that is predicted by the model is virtually absent (inset in Fig. 4). Such unusual dependence $H_{c2}(T)$ was explained in the frame of the 2-band model under assumption of two groups of carrier with essentially different Fermi velocities $v_F$ [24]. In this case, the HW model in the clean limit gives a positive curvature in the vicinity of $T_c$ and quasi-linear behavior at low temperatures. Roughly speaking, in the vicinity of $T_c$ the slope of $dH_{c2}/dT$ at $T \to T_c$ is controlled by light carriers with larger $T_c$ and smaller $H_{c2}(0)$ values, whereas at $T \to 0$, heavy carriers with a smaller $T_c$ and larger $H_{c2}(0)$. All these peculiarities must disappear upon transition to the dirty limit.

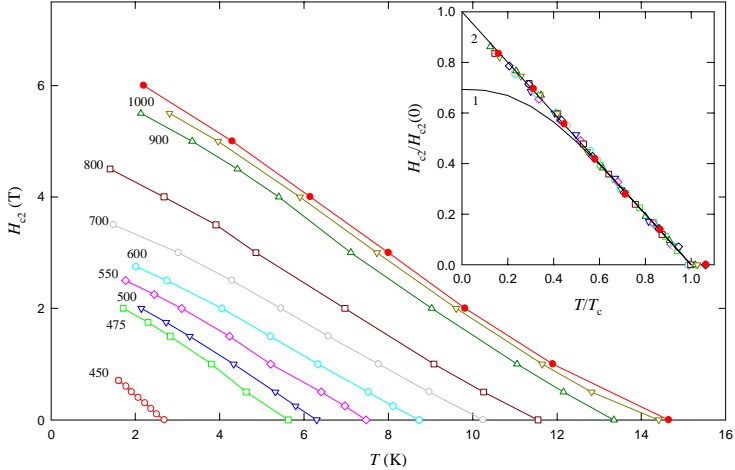

Fig. 4. Temperature dependences of the upper critical field $H_{c2}(T)$ for the sample of $YNi_2B_2C$; curve notation are identical to those in Fig. 1. In the inset there are shown curves in the reduced coordinates; black lines show theoretical curves in the frame of the HW model [23] in the dirty limit (1) and linear dependence (2).

Actually, for the sample of $YNi_2B_2C$ in more disordered states at $T_c < 13$ K, no positively curved portions are present (Fig. 4), however, the quasilinear dependence at low temperatures is retained. A very similar behavior of $H_{c2}/H_{c2}(0)$ versus $T_c/T_c(0)$ was observed also in the doped systems $Y(Ni_{1-x}Co_x)_2B_2C$ [16] and $Lu(Ni_{1-x}Co_x)_2B_2C$ [25].

Figure 5 presents the summed results of annealings of the irradiated $YNi_2B_2C$ sample. Here, the values of $T_c$ and $-dH_{c2}/dT$ are obtained via interpolation of the experimental curves $H_{c2}(T)$ with a linear dependence, with the exception of points for which the curvature of $H_{c2}(T)$ is positive in the vicinity $T_c$ for the sample in the state with $T_c > 13$ K, and $\rho_0$ and $a_2$ are obtained via interpolation of the experimental curves $H_{c2}(T)$ with a quadratic dependence $\rho(T) = \rho_0 + a_2 T^2$ in the range $T_c < T < 70$ K. Judging from the dependences $\rho_0(T_{ann})$, the recombination of radiation defects takes place in a very wide range of temperatures $100 - 1000\,^\circ C$, which is apparently traceable to a complexity of the crystalline structure and, consequently, variety of migration paths for defects. The fastest change of $\rho_0$ occurs in the range $400 - 500\,^\circ C$, where the SC state manifests itself for the first time as well.

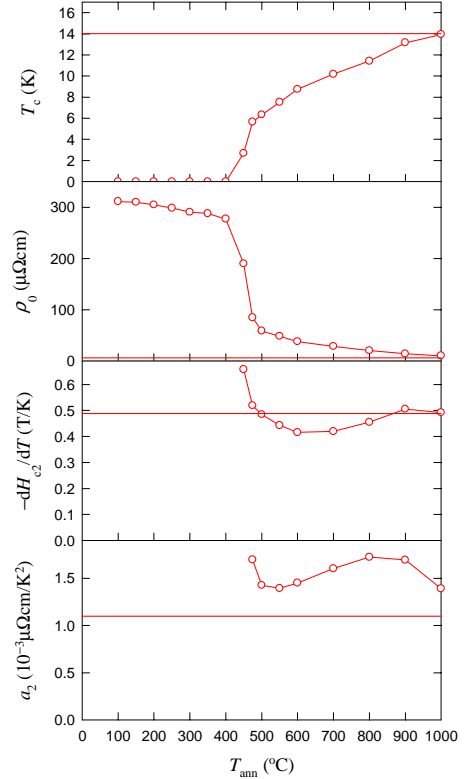

Fig. 5. Temperature of the SC transition $T_c$, residual resistivity $\rho_0$, slope of the second critical field $-dH_{c2}/dT$, and coefficient $a_2$ in the relationship $\rho(T) = \rho_0 + a_2 T^2$ for the irradiated sample of $YNi_2B_2C$ as a function of annealing temperature $T_{ann}$. Solid horizontal lines show the corresponding values for the sample in the initial state.

The $T_c$ versus $\rho_0$ curve for the irradiated sample of $YNi_2B_2C$ presents a rather unusual dependence (рис. 6). In the range 6 K < $T_c$ < 15 K the dependence $T_c(\rho_0)$ is approximately linear and is extrapolated to $\rho_0 \approx 90$ μΩcm at $T_c \to 0$. In the region $T_c < 6$ K the dependence $T_c(\rho_0)$ is also roughly linear with points of intersection $\rho_0 \approx 270$ μΩcm at $T_c = 0$ and $T_c \approx 8$ K at $\rho_0 = 0$. Such dependence can be interpreted as a suppression by disordering of two different SC gaps $\Delta_1$ and $\Delta_2$ ($\Delta_1 \sim 2\Delta_2$), though the rate of suppression of $\Delta_1$ is approximately 3 times as high as in the case of $\Delta_2$. Consequently, the value of $T_c$ to the left of the point of bending at $\rho_0 \approx 60$ μΩcm is controlled by the gap $\Delta_1$, whereas to the right, $\Delta_2$. Note a significantly faster decrease of $T_c$ upon doping in the system Lu(Ni$_{1-x}$Co$_x$)$_2$B$_2$C [14].

In case of the quasilinear dependence $H_{c2}(T)$, the slope of the upper critical field $-dH_{c2}/dT \approx H_{c2}(0)/T_c = \Phi_0/(2\pi T_c \xi^2)$, where the SC coherence length $\xi \approx \xi_0$, in the clean limit, $\xi_0 \gg l_{tr}$ and $\xi \approx (\xi_0 l_{tr})^{1/2}$, whereas in the dirty limit, $\xi_0 \ll l_{tr}$, $\xi_0 \approx (\hbar v_F)/(2\pi k_B T_c)$ (here we knowingly omit coefficients of the order of unit in view of a proximity of the following estimations).

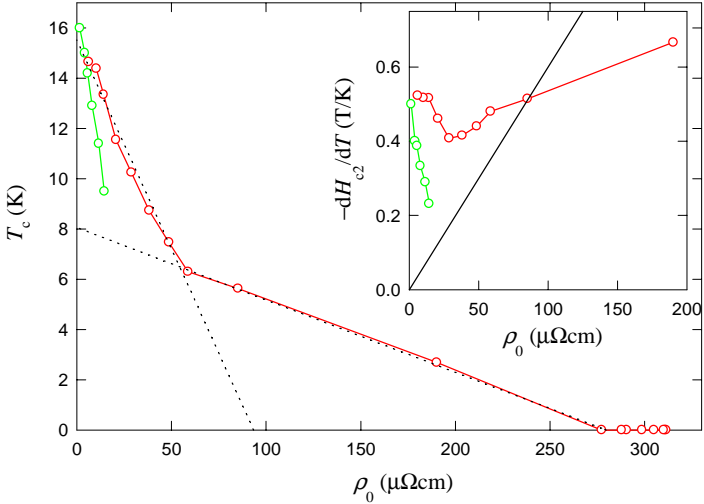

Fig. 6. Dependences of $T_c$ and $-dH_{c2}/dT$ (inset) on $\rho_0$ for YNi$_2$B$_2$C (○) and the doped compound Lu(Ni$_{1-x}$Co$_x$)$_2$B$_2$C (○) [14]. Dash-and-dot lines trace the experimental data; solid black line in the inset shows the expected behavior of $dH_{c2}/dT$ in the dirty limit, equality (1).

Thus, in the clean limit, $(-dH_{c2}/dT)_{clean} \sim T_c$, whereas in the dirty limit

$$(-dH_{c2}/dT)_{dirty} = (\Phi_0 e^2)/(\hbar \pi^3 k_B) \gamma_V \rho_0, \quad (1)$$

i.e., it is proportional to $\rho_0$ ($\gamma_V$ is the coefficient of electronic heat capacity per unit volume, $\gamma_V = 5.2 \cdot 10^2$ J/(m$^3$K$^2$) [3, 4]). The dependence of $-dH_{c2}/dT$ on $\rho_0$ (inset in Fig.6) qualitatively agrees with the expected behavior for a superconductor in an intermediated limit. The decrease of $-dH_{c2}/dT$ in the region of low $\rho_0$ values in the irradiated YNi$_2$B$_2$C (at $\rho_0 \leq 25$ μΩcm) and doped Lu(Ni$_{1-x}$Co$_x$)$_2$B$_2$C is related to the decrease of $T_c$ in the clean limit. Upon the transition to the dirty limit, $-dH_{c2}/dT$ starts linearly. In the intermediate limit one can write down $1/\xi^2 \approx 1/(\xi_0)^2 + 1/(\xi_0 l_{tr})$, or $-dH_{c2}/dT \approx$

$(-dH_{c2}/dT)_{clean} + (-dH_{c2}/dT)_{dirty} = c_1 T_c + c_2 \rho_0$, where the coefficients $c_1$ and $c_2$ depend only on the band parameters $v_F$ and $N(E_F)$. Assuming the band parameters constant, which agrees with the observed weak changes of the parameter $a_2$ (Fig. 2, 5), we obtain the linear dependence

$$(-dH_{c2}/dT)/T_c = c_1 + c_2(\rho_0/T_c), \quad (2)$$

which is shown with a straight line in Fig. 7. Qualitatively, the linear dependence reproduces itself quite well for the irradiated sample of YNi$_2$B$_2$C at $\rho_0 \leq 60$ μΩcm, the point of bend in Fig. 6; at $\rho_0 > 60$ μΩcm the experimental slope is 2-3 times as low. Such a behavior is likely to be connected with the different rates of suppression of gaps $\Delta_1$ and $\Delta_2$ upon disordering. Note that for the doped system Lu(Ni$_{1-x}$Co$_x$)$_2$B$_2$C (Fig. 7) there is observed a significant decline from the linear dependence (2), which obviously results from a significant decrease of $N(E_F)$.

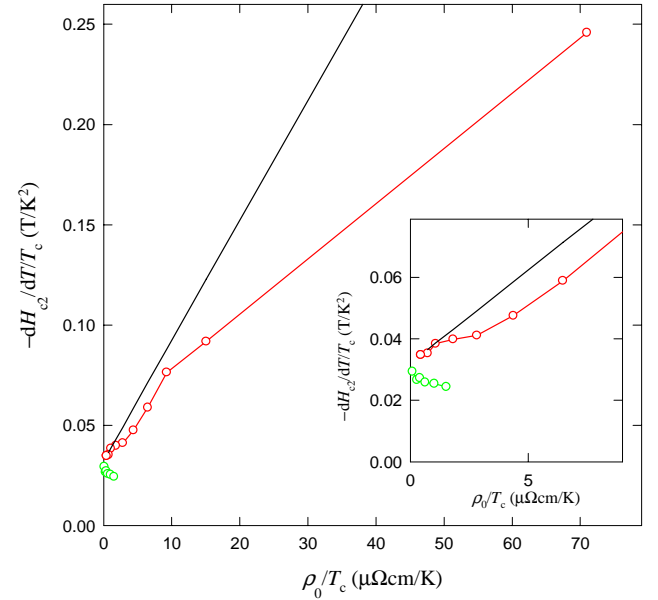

Fig. 7 Dependences of $(-dH_{c2}/dT)/T_c$ on $\rho_0/T_c$ for YNi$_2$B$_2$C (○) and doped Lu(Ni$_{1-x}$Co$_x$)$_2$B$_2$C compounds (○) [14]. Solid black line shows the expected behavior of $(-dH_{c2}/dT)/T_c$ in the intermediate limit, equality (2). In the inset, the curves are shown on a larger scale.

Thus, the observed suppression of superconductivity upon irradiation of YNi$_2$B$_2$C clearly points to an unconventional type of symmetry of the SC gaps, both $\Delta_1$ and $\Delta_2$, provided the density of states $N(E_F)$ changes insignificantly.

To compare the results with the theoretical models,

let us use the universal Abrikosov-Gor'kov equation, which describes the suppression of superconductivity by magnetic impurities in case of conventional $s$-pairing or nonmagnetic impurities (defects) in case of sign-changing gap ($d$- or $s^{\pm}$-pairing) [15, 26, 27]:

$$\ln(1/t) = \psi(g/t + 1/2) - \psi(1/2), \quad (3)$$

where $g = \hbar/(2\pi k_B T_c(0)\tau) = \xi_0(0)/l_{tr}$, $\psi$ is the digamma function, $t = T_c/T_c(0)$, $T_c(0)$, and $T_c$ are the temperatures of the SC transition in the initial and defect systems, $\tau$ is the time of electronic relaxation, $\xi_0(0) = (\hbar v_F)/(2\pi k_B T_c(0))$. Equation (3) describes the decrease of $T_c$ as a function of relaxation time $\tau$. The superconductivity becomes suppressed at $g > g_c = 0.28$. The dimensionless parameter $g$ can be constructed based on the experimental values of $g = (\hbar \rho_0)/(2\pi k_B T_c \mu_0 \lambda_c^2)$, where $\lambda$ is the SC penetration depth, and in this way, one can directly compare experimental data with the AG theory. However, such scheme works only in the case of one band (for example, $d$-pairing in the Cu-based HTSC). In the many-band case (for example, $s^{\pm}$-pairing in the Fe- based HTSC) in equation (3) there enters only interband scattering, which is rather difficult to separate out of the total contribution to the value $\rho_0$. It would be better to employ the relationship $g = \xi_0(0)/l_{tr}$, basing on equality (2), for the determination of the ratio $\xi_0(0)/l_{tr}$, since it can be expected that in this case the pairing and scattering are connected with one and the same interaction. Let us rewrite (2) in the form

$$(-dH_{c2}/dT)/T_c = \Phi_0/(2\pi(T_c)^2 \xi^2)$$
$$\approx \Phi_0/(2\pi(T_c(0))^2(\xi_0(0))^2)(1 + gT_c(0)/T_c).$$

Further, supposing the value of $\xi_0(0)$ and relationship $g \sim 1/l_{tr}$ being preserved in a defect system, the intercept in the limit $\rho_0/T_c \to 0$ gives the coherent length $\xi = \xi_0$ in the clean limit (Fig. 7). The double intercept on the experimental curve gives $gT_c(0)/T_c = 1$, which corresponds to $\rho_0 \approx 50$ μΩcm, $T_c \approx 7.5$ K, $g \approx 0.5$. Hence, we obtain an approximate relationship $g \approx 0.01\rho_0$. Then, the suppression of the gaps $\Delta_1$ and $\Delta_2$ upon disordering (Fig. 6) takes place at $g_1 \approx 0.9$ and $g_2 \approx 2.7$, respectively, which far exceed the AG value $g_c = 0.28$.

Another way of estimating the parameter $g$ is to specify $l_{tr}$ for a strongly disordered state as $l_{tr} \approx a_0 = (V_{cell})^{1/3} \approx 5$ Å for YNi$_2$B$_2$C ($V_{cell}$ is the unit-cell volume), provided, similarly, that the band parameters in a defect state remain unchanged. The coherency length (in the limit $\rho_0 \to 0$) $\xi_0(0) \approx (\Phi_0/(2\pi(-dH_{c2}/dT)))^{1/2} \approx 65$ Å [17], which gives $g = \xi_0(0)/l_{tr} \approx 13$ at $\rho_0 \approx 300$ μΩcm (Fig. 1). Hence, we obtain $g \approx 0.04\rho_0$ and, consequently, $g_1 \approx 3.6$ and $g_2 \approx 11$. These estimates are even more different from the AG model, which is in part accounted for by taking into account all (intra- and interband) scattering processes and this is what distinguishes this case from the previous one.

Thus, application of the AG equation results in a significant overestimation of the rate of decrease of $T_c$, just as in the case of many other unconventional superconductors such as Fe-based HTSC [28, 29] or Lu$_2$Fe$_3$Si$_5$ [30]. The reasons for these discrepancies can be connected just with the many-band nature of the superconductors. In this case, a relatively slower decrease of $T_c$ can be conditioned by a weak nonmagnetic scattering between those portions of the Fermi surface which take part in the SC pairing. For instance, in the case of $s^{\pm}$-model [31], $T_c$ does not change at all if the interband scattering is absolutely absent and, consequently, if the interband, scattering is weak (compared to the intraband ones), the degradation of superconductivity must take place at essentially higher values of $\rho_0$ than it follows from the one-band AG model. Anyhow, the complete degradation of superconductivity upon irradiation of YNi$_2$B$_2$C cannot obviously match the electron-phonon mechanism of superconductivity, which suggests that the most probable reason for a notable decrease of $T_c$ be the decrease of the density of electron states $N(E_F)$.

In conclusion, our results indicate a fast decrease of the temperature of the SC transition, $T_c$, in the polycrystalline sample of YNi$_2$B$_2$C upon neutron irradiation which effectively generates nonmagnetic scattering centers. The analysis performed for the coefficient $a_2$ in the electrical resistivity $\rho(T) = \rho_0 + a_2 T^2$ and the slope of the upper critical field $-dH_{c2}/dT$ as functions of $\rho_0$ and $T_c$ shows a relatively weak change of the density of electron states $N(E_F)$. Since it is the decrease of $N(E_F)$ that can be the reason for a decrease of $T_c$ in the systems with the conventional SC pairing of the $s$-type (electron-phonon interaction), the complete degradation of superconductivity in YNi$_2$B$_2$C upon irradiation indicates that this compound can be referred to as an unconventional superconductor with the sign-changing SC gap function, for which magnetic excitations serve as quasiparticles resulting in the Cooper-type pairing of electrons. The dependence of $T_c$ on $\rho_0$ can be inter-preted as a suppression of two different SC gaps $\Delta_1$ and $\Delta_2$ ($\Delta_1 \sim 2\Delta_2$) induced by defects, with the rate of suppression of $\Delta_1$ exceeding that of $\Delta_2$ by a factor of ~ 3.

The work was carried out according to the Plan of RAS (theme "Pulse" № 01.2.006 13394) with a partial Support of the Program for Basic Research of the Presidium of RAS "Quantum mesoscopic and disordered structures" (project № 12-Р-12-П-2-1018 of the Ural Branch, RAS), Russian Foundation for on Basic Research (project № 11-02-00224), and State Contract of the Ministry of Education and Science (project № 14.518.11.7020